\documentclass[twocolumn,showpacs,aps,prl,superscriptaddress]{revtex4}
\usepackage{graphicx} 
\usepackage{dcolumn}
\usepackage{epsfig}    
\usepackage{amsmath}

\newcommand{\BaBarPubYear}    {10}
\newcommand{\BaBarPubNumber}  {022}
\newcommand{\SLACPubNumber} {14406}
\newcommand{\LANLNumber} {1103.3971}

\newcommand{\BaBarType}      {PUB}  

\input babarsym   
\newcommand{\calB}{\mbox{${\cal B}$}}
\newcommand{\calH}{\mbox{${\cal H}$}}

\def\etal{{\em et al.}}

\newcommand{\km}{\mbox{$K^-$}}

\newcommand{\kp}{\mbox{$K^+$}}

\def\babar{{\em B}{\footnotesize\em A}{\em B}{\footnotesize\em AR}}

\def\gevc{\mbox{${\mathrm{GeV}}/c\ $}}
\def\gevcc{\mbox{${\mathrm{GeV}}/c^2\ $}}
\def\mev{\mbox{${\mathrm{MeV}}\ $}}
\def\kev{\mbox{${\mathrm{keV}}\ $}}

\def\Bz{\mbox{$\overline {B^{0}}\ $}}

\def\BB{\mbox{$\B^{+}\  \B^{-}\ $}}

\def\Bz{\mbox{$B^0\ $}}
\def\Bzb{\mbox{$\overline{B}^0\ $}}
\def\beq{\begin{equation}}
\def\eeq{\end{equation}}
\def\bef{\begin{figure}}
\def\edf{\end{figure}}
\def\ben{\begin{enumerate}}
\def\een{\end{enumerate}}
\def\bear{\begin{array}}
\def\enar{\end{array}}
\def\beqa{\begin{eqnarray}}
\def\eeqa{\end{eqnarray}}

\newcommand {\dEdx}     {\mbox{$\mathrm{d}E/\mathrm{d}x$}}

\def\to{\mbox{$\rightarrow$}}

\def\gevc{\mbox{${\mathrm{GeV}}/c$}}
\def\gevcc{\mbox{${\mathrm{GeV}}/c^2$}}
\def\mev{\mbox{${\mathrm{MeV}}$}}
\def\mevcc{\mbox{${\mathrm{MeV}}/c^2$}}

\def\invfb{\mbox{${\mathrm{fb}^{-1}}$}}
\def\FourS{\mbox{$\Upsilon{\mathrm( 4S)}$}}

\def\Bz{\mbox{${B^{0}}$}}
\def\Bzb{\mbox{$\overline B^{0}$}}
\def\BB{\mbox{$B\overline B$}}
\def\BpBm{\mbox{$B^{+} B^{-}$}}

\def\piz{\mbox{${\pi^{0}}$}}
\def\pip{\mbox{${\pi^{+}}$}}
\def\pim{\mbox{${\pi^{-}}$}}

\def\pimp{\mbox{${\pi^{\mp}}$}}

\def\Kp{\mbox{${K^{+}}$}}

\def\pipi     {\ensuremath{\pi\pi}}

\def\kk     {\ensuremath{KK}}

\newcommand{\etacts}{\ensuremath{\eta_{c}(2S)}}

\newcommand{\psit}{\ensuremath{\psi(2S)}}

\newcommand{\chicz}{\ensuremath{\chi_{c0}(1P)}}
\newcommand{\chict}{\ensuremath{\chi_{c2}(1P)}}
\newcommand{\chics}{\ensuremath{\chi_{c0,2}(1P)}}
\newcommand{\ks}{\ensuremath{\KS K^{\pm}\pi^{\mp}}}
\renewcommand{\kk}{\ensuremath{\kp\km\pip\pim\piz}}
\renewcommand{\mm}{\ensuremath{M^2_{\mathrm{miss}}}}

\renewcommand{\gg}{\ensuremath{\gamma\gamma}}
\renewcommand{\etac}{\ensuremath{\eta_c(1S)}}

\begin{document}

\begin{flushleft}
~\\
~\\
\end{flushleft}

\begin{flushright}
~\\
~\\
\babar-\BaBarType-\BaBarPubYear/\BaBarPubNumber \\
SLAC-\BaBarType-\SLACPubNumber \\
hep-ex/\LANLNumber
\end{flushright}

\title{
 \large \bf\boldmath Observation of $\eta_c(1S)$ and $\eta_c(2S)$
 decays to \kk\ in two-photon interactions
}

%
\author{P.~del~Amo~Sanchez}
\author{J.~P.~Lees}
\author{V.~Poireau}
\author{E.~Prencipe}
\author{V.~Tisserand}
\affiliation{Laboratoire d'Annecy-le-Vieux de Physique des Particules (LAPP), Universit\'e de Savoie, CNRS/IN2P3,  F-74941 Annecy-Le-Vieux, France}
\author{J.~Garra~Tico}
\author{E.~Grauges}
\affiliation{Universitat de Barcelona, Facultat de Fisica, Departament ECM, E-08028 Barcelona, Spain }
\author{M.~Martinelli$^{ab}$}
\author{A.~Palano$^{ab}$ }
\author{M.~Pappagallo$^{ab}$ }
\affiliation{INFN Sezione di Bari$^{a}$; Dipartimento di Fisica, Universit\`a di Bari$^{b}$, I-70126 Bari, Italy }
% ...
\collaboration{The \babar\ Collaboration}
\noaffiliation

\date{\today}

\begin{abstract}
We study the processes $\gg\to\ks$ and $\gg\to\kk$ using a data sample
of 519.2~\invfb\
recorded by the \babar\ detector at the PEP-II asymmetric-energy
\epem\ collider at center-of-mass energies near the
$\Upsilon(nS)$ ($n = 2,3,4$) resonances.
We observe the \etac, \chicz\  and
$\eta_c(2S)$ resonances produced in two-photon interactions and
decaying to \kk, with significances of  $18.1$, $5.4$ and
$5.3$ standard deviations (including 
systematic errors), respectively, and report $4.0\sigma$ evidence of
the 
\chict\ decay to this final state.
We measure the $\eta_c(2S)$
mass and width in \ks\ decays, and obtain the values
$m(\eta_c(2S))=3638.5 \pm 1.5 \pm 0.8~\mevcc$ and 
$\Gamma(\eta_c(2S)) = 13.4\pm4.6\pm 3.2~\mev$, where
the first uncertainty is statistical and the second is systematic.
We measure the two-photon width times branching
fraction for the reported resonance signals, and search for
the $\chi_{c2}(2P)$ resonance, but no significant signal
is observed. 
\end{abstract}

\maketitle

The first radial excitation \etacts\ of the \etac\ charmonium ground state
was observed at
$B$-factories~\cite{etacts,CLEO,etactsBaBar,etactsBaBarIncl}. The only
observed exclusive decay of this state to date is to
$K\overline{K}\pi$~\cite{PDG}.  
Decays to $p\bar{p}$ and $h^{+}h^{-}h^{\prime+}h^{\prime-}$, 
with $h^{(\prime)}=K, \pi$, have been observed for
the \etac~\cite{PDG}, but not for the \etacts~\cite{Uehara,Ambrogiani}.
Precise determination of the \etacts\ mass may discriminate
among models that predict the \psit-\etacts\ mass
splitting~\cite{massSplit}. 

After the discovery of the $X(3872)$ state~\cite{BelleX} and its
confirmation by different experiments~\cite{Xconfirmation},
charmonium spectroscopy above the open-charm threshold received renewed
attention. Many new states have been established to
date~\cite{newStates,BelleZ,BaBarZ}. The $Z(3930)$ 
resonance was discovered by Belle in the $\gg\to D\overline{D}$
process~\cite{BelleZ} and subsequently confirmed by
\babar~\cite{BaBarZ}. Its
interpretation as the $\chi_{c2}(2P)$, the first radial excitation of the
$\phantom{}^3P_2$ charmonium ground state, is commonly
accepted~\cite{PDG}.  

In this paper we study charmonium
resonances produced in the two-photon process $\epem\to\gg\epem\to
f\epem$, where $f$ denotes the $\ks$ or \kk\ final state.
Two-photon events where the interacting photons are not quasi-real are
strongly suppressed by the selection 
criteria described below. This implies that the allowed $J^{PC}$ values of
the initial state are $0^{\pm+}$, $2^{\pm+}$, $4^{\pm+}$, ...;
$3^{++}$, $5^{++}$, ...~\cite{Yang}.        
Angular momentum conservation, parity conservation, and charge conjugation
invariance, then imply that these quantum numbers apply to
the final states $f$ also, except that the \ks\ state cannot have $J^P
= 0^+$. 

The results presented here are based on data collected
with the \babar\ detector
at the PEP-II asymmetric-energy $e^+e^-$ collider,
corresponding to 
an integrated luminosity of 519.2~\invfb, recorded at center-of-mass (CM)
energies near the $\Upsilon (nS)$ ($n=2,3,4$) resonances.

The \babar\ detector is described in detail elsewhere~\cite{BABARNIM}.
Charged-particles resulting from the interaction are detected, and their
momenta are measured, by a combination of five layers of double-sided
silicon microstrip detectors and a 40-layer drift chamber.
Both systems operate in the 1.5~T magnetic field of a superconducting
solenoid. Photons and electrons are identified in a CsI(Tl) crystal
electromagnetic calorimeter. Charged-particle
identification (PID) is provided by the specific energy loss (\dEdx) in
the tracking devices, and by an internally reflecting, ring-imaging
Cherenkov detector.
Samples of Monte Carlo (MC) simulated events~\cite{geant}, which are
more than 10 times larger than the corresponding data samples, are
used to study signals and 
backgrounds. Two-photon events are generated using the GamGam
generator~\cite{BaBarZ}.

Neutral pions and kaons are reconstructed through the
decays $\piz\to\gamma\gamma$ and $\KS\to\pip\pim$.
Photons from \piz\ decays are required to have laboratory energy
larger than 30~\mev. We require the invariant mass of a \piz\ (\KS)
candidate to be in the range [100--160] ([470--520])~\mevcc.
Neutral pions reconstructed with these criteria are used to veto events
with multiple \piz\ mesons, as described below.
For the \kk\ mode, we refine the selection of the \piz\ by requiring the
laboratory energy of the lower-energy 
photon from the signal \piz\ decay to be larger than
$50~\mev$. Furthermore, we require $|\cos\calH_{\piz}|<0.95$, where
$\calH_{\piz}$ is the angle between the signal \piz\ flight
direction in the laboratory frame and the direction of one of its
daughters in the \piz\ rest frame.  These requirements are optimized by
maximizing $S/\sqrt{S+B}$, where $S$ is the number of MC signal
events with a well-reconstructed \piz, and $B$ is the
combinatorial background in the signal region.
Primary charged-particle tracks are
required to satisfy PID requirements consistent with a kaon or 
pion hypothesis.
A candidate event is 
constructed by fitting the \piz\ (\KS) candidate and four (two)
charged-particle tracks of zero net charge coming from the interaction
region to a common vertex. 
In this fit the \piz\ and \KS\ masses are constrained to their
nominal values~\cite{PDG}. We require the vertex fit probability of
the charmonium candidate to be larger than 0.1\%. 
The outgoing $e^{\pm}$ are not detected.

\begin{figure}[!h]
   \begin{center}
     \includegraphics[scale=0.45]{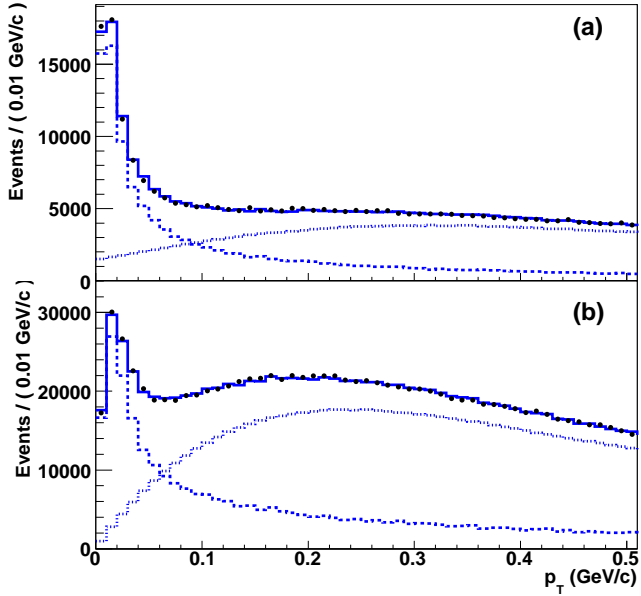}
     \caption{The \pt\ distributions for selected (a) \ks\ and (b) \kk\
       candidates (data points). The solid histogram represent the
       result of a fit to the sum of the simulated signal (dashed) and
       background (dotted) contributions.}
   \label{fig:pt}
   \end{center}
 \end{figure}

Background arises mainly from random combinations of particles from
\epem\ annihilation, other two-photon collisions, and initial state
radiation (ISR) 
processes. To suppress 
these backgrounds, we 
require that each event have exactly four charged-particle tracks. 
The candidate event is rejected if the number of additional reconstructed
photons is larger than 6 (5) for \kk\ (\ks). 
Similarly, the event is rejected if the number of additional
reconstructed \piz's is larger than 1 (3) for a \kk\
(\ks) candidate event. 
We discriminate against ISR background by
requiring
$\mm=(p_{\epem}-p_{\mathrm{rec}})^2>2~(\gevcc)^2$, where
$p_{\epem}$ ($p_{\mathrm{rec}}$) is the four 
momentum of the initial state (reconstructed final state). 
The effect of this requirement on the signal efficiency is studied
using a \kp\km\pip\pim\ control sample that contains large \etac,
\jpsi, and \chics\ 
signals. 
Two-photon events are expected to have low transverse momentum (\pt)
with respect to the collision axis.  
In Fig.~\ref{fig:pt}, we show the \pt\ distribution for 
selected candidates with the above requirements. The distribution
is fitted 
with the signal \pt\ shape obtained from MC simulation plus a
combinatorial background component, modeled using a sixth-order
polynomial function. We require $\pt<0.15~\gevc$. 

\begin{figure}[!h]
   \begin{center}
     \includegraphics[scale=0.48]{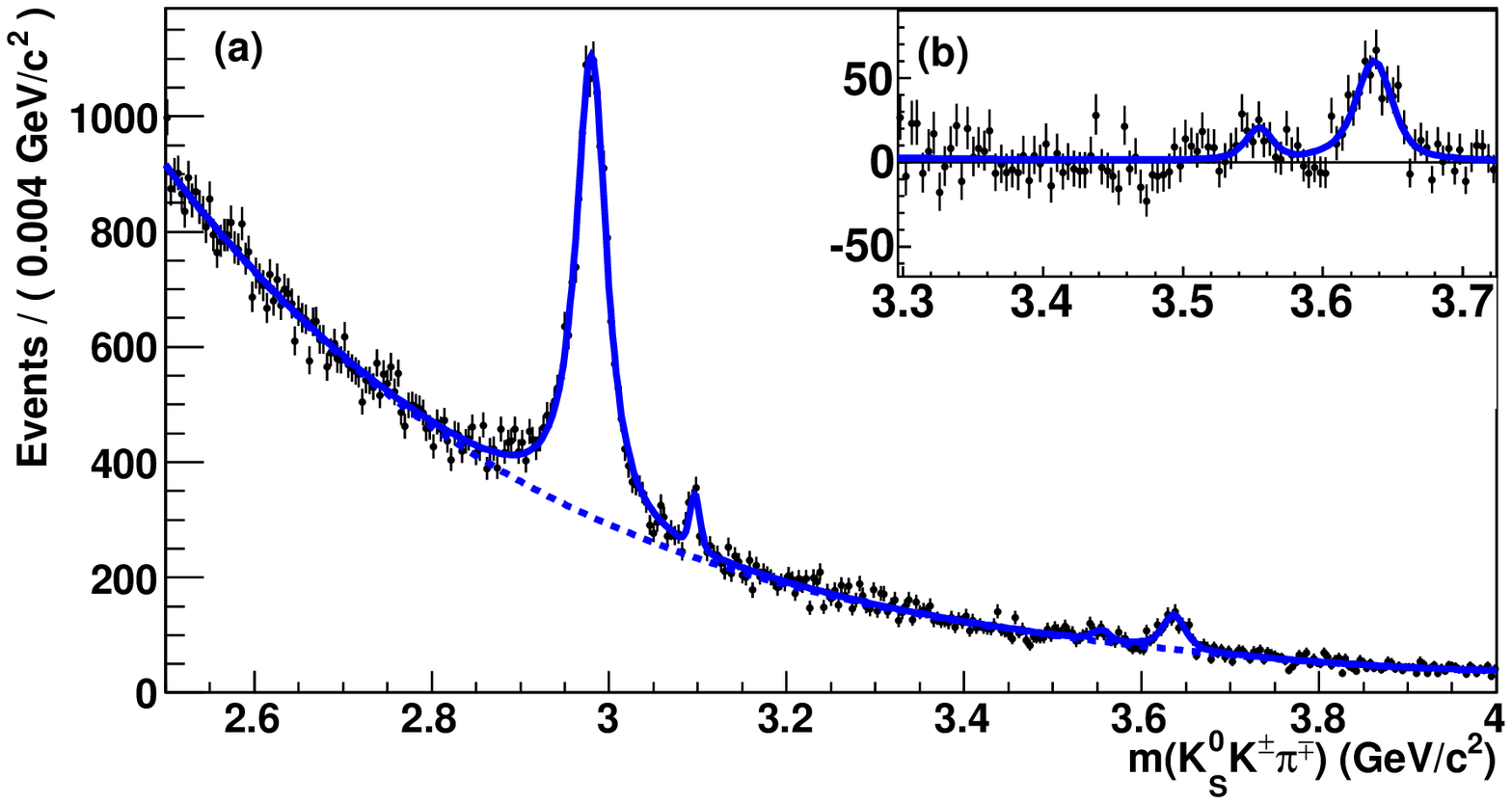}
     \includegraphics[scale=0.48]{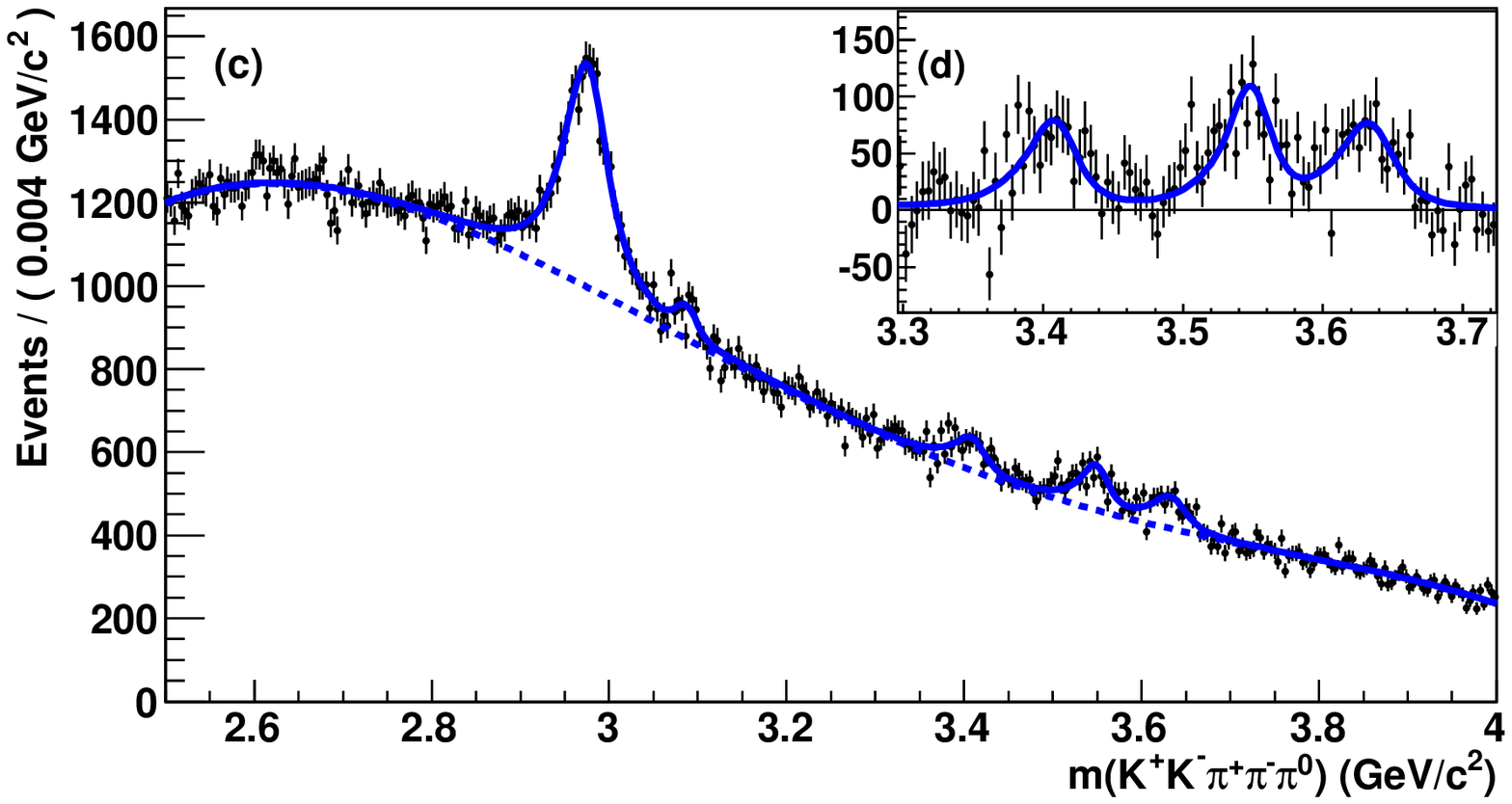}
     \caption{Fit to (a) the \ks\ and (c) the \kk\ mass spectrum. The
       solid curves represent the total fit functions and the dashed curves
       show the combinatorial background contributions. The
       background-subtracted distributions are shown in (b) and (d),
       where the 
       solid curves indicate the signal components. 
     }
     \label{fig:fit}
   \end{center}
 \end{figure}

The average number of surviving candidates per selected event is 1.003
(1.09) for the \ks\ (\kk) final state. 
Candidates that are rejected by a possible best-candidate selection do
not lead to any peaking structures in the mass spectra, and so no
best-candidate selection is performed. 
The $\KS\kp\pim$ and $\kp\km\pip\pim\piz$ mass spectra are shown in
Fig.~\ref{fig:fit}.  We observe prominent peaks at the position of the
\etac\ resonance. We also observe signals at the positions of the
\jpsi, \chicz, \chict, and \etacts\ states. 

The resonance signal yields and the mass and width of the \etac\ and
\etacts\ are extracted using a binned,
extended maximum likelihood fit to the invariant mass
distributions. The bin width is 4~\mevcc. 
In the likelihood function, several components are present:
\etac, \chicz, \chict, and \etacts\ signal, 
combinatorial background, and \jpsi\ ISR background.
The \chicz\ component is not present in the fit to the \ks\
invariant mass spectrum, since $J^P = 0^+$ is forbidden for this final
state.  

We parameterize each signal PDF as a convolution of a
non-relativistic Breit-Wigner and the detector resolution
function. The \jpsi\ ISR background is parameterized with a
Gaussian shape, and the combinatorial background PDF is a fourth-order
polynomial.
The free parameters of the fit are the yields of the resonances and
the background,
the peak masses and widths of the \etac\ and \etacts\ signals, the
width of the 
Gaussian describing the \jpsi\ ISR background, and the background
shape parameters.
The mass and width of the $\chi_{c0,2}(1P)$
states (and the mass of 
the \jpsi\ in the \ks\ channel), are fixed to their nominal
values~\cite{PDG}. 
For the \kk\ channel, the \etacts\ width is fixed to the value found
in the \ks\ channel. 
%

We define a MC event as ``MC-Truth'' (MCT) if the
reconstructed decay chain matches the generated one.  
We use MCT signal and MCT ISR-background events to determine the detector
mass resolution function. This function is described by the
sum of a Gaussian plus power-law tails~\cite{BBW}.
The width of the resolution function at half-maximum for the \etac\ is
8.1 (11.8)~\mevcc\ in the \ks\ (\kk) 
decay mode.
For the \etacts\ decay it is 10.6 (13.1)~\mevcc\ in the \ks\ (\kk)
decay mode. 
The parameter values for the resolution functions,  are fixed to their
MC values in the fit. 
%

Fit results are reported in Table~\ref{tab:results} and
shown in Fig.~\ref{fig:fit}. We correct the fitted \etac\ yields by
subtracting 
the number of peaking-background events originating from the
$\jpsi\to\gamma\etac$ decay, 
estimated below.  The statistical significances of the signal yields are
computed from the ratio of the number of observed events to the sum in
quadrature of the statistical and systematic uncertainties.
The $\chi^2/ndf$ of the fit is 1.07 (1.03), where $ndf$ is the number of
degrees of freedom, which is 361 (360) for the fit to \ks\ (\kk).

To search for the $\chi_{c2}(2P)$, we add to the fit described above a signal
component with the mass and width fixed to the values reported
in Ref.~\cite{BaBarZ}. No significant changes are observed in the fit
results. 
\begin{table*}[!htb]
  \centering
\caption{Extraction of event yields and mass and width of the \etac\
  and \etacts\  resonances: average signal efficiency for phase-space
  MCT events, corrected signal yield with statistical and systematic
  uncertainties, number of peaking-background events estimated with
  the \pt\ fit ($N_{\mathrm{peak}}$), number of peaking-background
  events from \jpsi\ and $\psi(2S)$ radiative decays ($N_{\psi}$), 
  significance (including systematic
  uncertainty), corrected mass, and fitted width for each decay mode.
  We do not report $N_{\psi}$ for modes where it is negligible.} 
  \begin{tabular}{cccccccc}
    \hline\hline
    Decay & Efficiency       & Corrected    & $N_{\mathrm{peak}}$ & $N_{\psi}$
    &Significance & Corrected  & Fitted \\ 
    Mode & (\%)& Yield (Evts.)& (Evts.)&(Evts.)&($\sigma$) & Mass
    (\mevcc) & Width (\mev) \\
    \hline
\hline
$\etac\to\ks$ & 10.7 & $12096\pm235\pm274$& $189 \pm 18$& $214\pm82$ &
33.5 & $2982.5 \pm 0.4\pm1.4$& $32.1 \pm 1.1\pm1.3$\\ 

$\chict\to\ks$ & 13.1 & $ 126\pm37\pm14$& $-45\pm 11$& -- &3.2&  3556.2 
(fixed) & 2 (fixed)\\ 

$\etacts\to\ks$ & 13.3 & $624\pm72\pm34$& $25\pm5$& -- &7.8 &
$3638.5\pm1.5\pm0.8$&$13.4\pm4.6\pm3.2$\\  
\hline
$\etac\to\kk$ & 4.2 &
$11132\pm430\pm442$& $118\pm 32$& $26\pm9$ &18.1& 
$2984.5 \pm 0.8\pm3.1$& $36.2\pm2.8\pm 3.0$\\ 

$\chicz\to\kk$ & 5.6 & $1094\pm143\pm 143$& $-39\pm 19$& $75\pm21$ &5.4 &
3415.8 (fixed) & 10.2 (fixed)\\ 

$\chict\to\kk$ & 5.8 &
$1250\pm118\pm290$& $14\pm 24$& $233\pm73$&4.0 &3556.2 (fixed) &
2 (fixed)\\ 

$\etacts\to\kk$ &5.9 &
$1201\pm133\pm185$& $-46\pm 17$&--&5.3& 
 $ 3640.5\pm3.2\pm2.5$
& 13.4 (fixed)\\
\hline\hline
\end{tabular}
\label{tab:results}
\end{table*}
%
Several processes, including ISR, continuum \epem\ annihilation, and
two-photon events with a final state different from the one studied,
may produce irreducible-peaking-background events, containing real
\etac, \etacts, \chicz\ or \chict.
Well-reconstructed signal and \jpsi\ ISR background are expected to 
peak at $\pt\sim0~\gevc$. Final states with similar masses are
expected to have similar \pt\ distributions.
Non-ISR background events mainly originate from events with a number
of particles in the final state larger than the one in signal events.
Such extra particles are lost in the  reconstruction. Thus, non-ISR
background events are expected to have a nearly flat \pt\
distribution, as observed in MC simulation. 
 
To estimate the number of such events, we fit the invariant mass
distribution in intervals of \pt, thus obtaining the signal yield 
for each resonance as a function of \pt. The signal yield distribution
is then fitted using the signal \pt\ shape from MCT events
plus a flat background. 
The yield of peaking-background events originating from
$\psi$ radiative decays ($\psi=\jpsi,\psit$) is
estimated using the number of $\psi$ 
events fitted in data, and the knowledge of branching
fractions~\cite{PDG} and MC reconstruction efficiencies for the
different decays involved. 
The number of peaking-background events for
each resonance is reported in Table~\ref{tab:results}. 
The value of $\calB(\chi_{c0,2}\to\kk)$, which is needed to estimate
the number of peaking-background events from
$\psi(2S)\to\gamma\chi_{c0,2}(1P)$ decays, is obtained using the
results reported in this paper and the world-average values of
$\Gamma_{\gg}(\chi_{c0,2})$~\cite{PDG}.
We obtain $\calB(\chicz\to\kk) = (1.14\pm0.27)\%$, and 
$\calB(\chict\to\kk) = (1.30\pm0.36)\%$, where statistical and
systematic errors have been summed in quadrature. The value of
$\calB(\chict\to\kk)$ is in agreement with a 
preliminary result reported by CLEO~\cite{CLEOprel}.
The number of peaking background events from $\psi$ radiative decays
for \etacts\ and $\chict\to\ks$ (denoted by ``--'' in
Table~\ref{tab:results}) is negligible.

\begin{figure}[!htb]
   \begin{center}
     \includegraphics[scale=0.45]{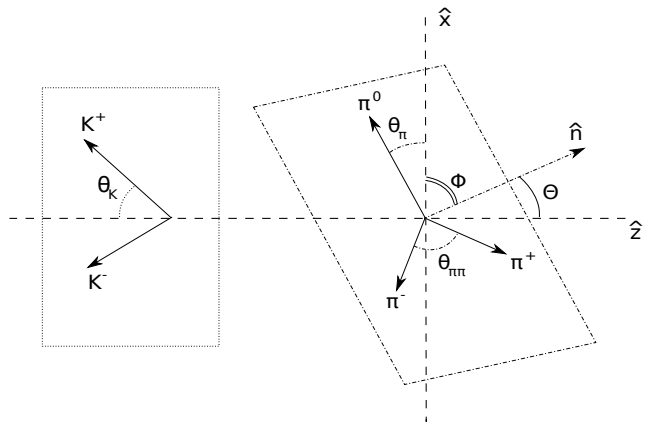}
     \caption{Angles used to describe the \kk\ decay kinematics}
     \label{fig:effvars}
   \end{center}
 \end{figure}

The ratios of the branching fractions of the two modes are obtained
from 
\begin{equation}
\frac{\calB(\eta_c(nS)\to\kk)}{\calB(\eta_c(nS)\to\ks)} = 
\frac{N^{\eta_c(nS)}_{KK3\pi}}{N^{\eta_c(nS)}_{\KS
    K\pi}}\cdot\frac{\epsilon^{\eta_c(nS)}_{\KS K\pi}}
{\epsilon^{\eta_c(nS)}_{KK3\pi}},
\label{eq:bf}
\end{equation}
where $\eta_c(nS)$ denotes \etac, \etacts; $N^{\eta_c(nS)}_{KK3\pi}$ and
$N^{\eta_c(nS)}_{K^{0}_{S}K\pi}$ ($\epsilon^{\eta_c(nS)}_{KK3\pi}$ and
$\epsilon^{\eta_c(nS)}_{K^{0}_{S}K\pi}$) represent the
peaking-background-subtracted 
$\eta_c(nS)$ yield (the efficiency) for the \kk\ and \ks
channels, respectively.  
The efficiencies are parameterized using MCT events.
The \ks\ efficiency is parameterized as a
two-dimensional histogram of 
the invariant $K\pi$ mass versus 
the angle between the direction of the \kp\ in the $K\pi$ rest frame
and that of the 
$K\pi$ system in the \ks\ reference frame.
The \kk\ efficiency is parameterized as a function of
the $K^+K^-$, $\pip\pim$, and
$\pip\pim\piz$ ($3\pi$) masses, and the five angular variables,
$\cos\theta_{K}$, $\cos\Theta$, $\Phi$,  $\cos\theta_{\pi\pi}$, and
$\theta_{\pi}$, as defined in Fig.~\ref{fig:effvars};
$\theta_{K}$ is the angle between the \kp\ and the $3\pi$ recoil
direction in the $K^+K^-$ rest frame. The angles $\Theta$ and $\Phi$ describe
the orientation of  the normal $\hat{n}$ to the $3\pi$ decay plane
with respect to the $K^+K^-$ recoil direction in the $3\pi$ rest frame; 
$\theta_{\pi}$ is the angle describing a 
rotation of the $3\pi$ system about its decay plane normal;
$\theta_{\pi\pi}$ is the angle between the \pip\ and \pim\
directions in the $3\pi$ reference frame. 
The correlations between $\cos\theta_K$,
$\Theta$, $\Phi$, and $\theta_{\pi}$ and the invariant masses are negligible.
The correlation between $\cos\theta_{\pi\pi}$ and $m_{\pi\pi}$ is
-0.70 and is not considered in the efficiency parameterization. 
Neglecting such a correlation introduces a change in the efficiency
of 1.4\% (1.1\%) for the \etac\ (\etacts), which is taken as a systematic
uncertainty. 
The efficiency dependence on $\cos\theta_K$, $\cos\theta_{\pi\pi}$,
 and $\Phi$ ($\cos\Theta$ and $\theta_\pi$) is parameterized using
uncorrelated fourth(second)-order polynomial shapes. 
A three-dimensional histogram is used to parameterize the dependence
on the invariant masses. 
The efficiency is calculated as the ratio of the number of MCT events 
surviving the selection to the number of generated events in each bin,
in both channels.
We assign null efficiency to bins with less than 10 reconstructed
events. The fraction of data falling in these bins is 0.5\% (3.0\%) 
in the \ks\ (\kk) channel. We assign a systematic uncertainty to cover
this effect.
The average efficiency $\overline{\epsilon}$ for each decay, computed
using flat phase-space simulation, is reported in
Table~\ref{tab:results}. 

The ratio $N^{\eta}_{f}/\epsilon^{\eta}_{f}$ of Eq.~(\ref{eq:bf}) is
equal to
$N^{\eta}_{f}/(\epsilon^{*\eta}_{f}\times\overline{\epsilon}^{\eta}_{f})$,
where we have defined $\epsilon^{*\eta}_{f}=
\epsilon^{\eta}_{f}/\overline{\epsilon}^{\eta}_{f}$.
The value of $N^{\eta}_{f}/\epsilon^{*\eta}_{f}$ is extracted from an
unbinned maximum  
likelihood  fit to the \ks\ and \kk\  invariant mass distributions,
where each event is weighted by the inverse of $\epsilon^{*\eta}_{f}$.
We use $\epsilon^{*\eta}_{f}$ instead of $\epsilon^{\eta}_{f}$ to
weight the events since weights far from one might result in incorrect
errors for the signal yield obtained in the maximum likelihood
fit~\cite{frodesen}. 
Since the kinematics of peaking-background events are similar to those of
the signal, we assume the signal to peaking-background ratio to be
unaffected by the weighting  
technique.
The fit is performed independently in the $\etac$ ($[2.5,3.3]~\gevcc$)
and $\etacts$ ($[3.2,3.9]~\gevcc$) mass regions.  
The mass and width for each signal PDF are fixed to
the values reported in Table~\ref{tab:results}. 
The free parameters of the fit are the yields of the background and
the signal resonances, the mean and the width of the 
Gaussian describing the \jpsi\ background, and the background shape
parameters. 
We compute a $\chi^2$ using the total fit function and the binned \ks\
(\kk) mass distribution obtained after weighting. The values of $\chi^2/ndf$
are 1.16 (1.15) and 1.20 (1.00) in the $\etac$ and $\etacts$ mass
regions, in the \ks\ (\kk) channel.    

%

Several sources contribute to systematic uncertainties on the
resonance yields and parameters.
Systematic uncertainties due to PDF parameterization and fixed
parameters in the fit are estimated to be the sum in quadrature of the
changes observed when repeating the fit after varying the
fixed parameters by $\pm1$ standard deviation ($\sigma$). The uncertainty
associated with the
peaking background is taken to be $\sqrt{(\max[0,N_{\rm peak}])^2 +
  \sigma_{N_{\rm peak}}^2}$, where $N_{\rm peak}$ is the estimated number
of peaking-background events reported in
Table~\ref{tab:results}, and $\sigma_{N_{\rm peak}}$ is
its uncertainty. The systematic errors on the $\chi_{c0,2}(1P)$
yields are taken to be $\sqrt{(\max[0,N_{\rm peak}])^2 +
  \sigma_{N_{\rm peak}}^2 + 
N_{\psi}^2+\sigma_{N_{\psi}}^2}$, where $N_\psi$ is the number of
peaking-background events from the $\psi(2S)\to\gamma\chi_{c0,2}(1P)$
processes. The uncertainty on $N_{\rm peak}$ due to differences in
signal and ISR background \pt\ distribution is estimated by adding an ISR
background component to the fit to the \pt\ yield distribution described
above. The ISR background \pt\ shape is taken from MC simulation and
its yield is fixed to $N_{\psi}$. This uncertainty is found to be
negligible. 
We take the systematic error due to the
$\jpsi\to\gamma\etac$ peaking-background subtraction to be the
uncertainty on the estimated number of events originating from this
process. 
We assign an uncertainty due to the background
shape, taking the changes in results observed when using a sixth-order
polynomial as the background PDF in the fit.

An ISR-enriched sample is obtained by reversing the \mm\ selection
criterion.  
The ISR-enriched sample is fitted to obtain the shift $\Delta M$
between the measured and the nominal \jpsi\
mass~\cite{PDG}, and the difference in mass resolution between MC
and data. 
The corrected masses in Table~\ref{tab:results} are $m^{corr}=
m^{fit}-\Delta M$, where $m^{fit}$ is the mass determined by the fit. 
The mass shift is $-0.5\pm0.2$~\mevcc\ in \ks\ and
$-1.1\pm0.8$~\mevcc\ in \kk. We assign the statistical error on
$\Delta M$ as a systematic uncertainty on $m^{corr}$. 
The difference in mass resolution is
$(24\pm5)\%$ in \ks\ and $(9\pm5)\%$ in \kk.
We take the difference in fit results observed when including
this correction in the \etac, \chicz, \chict, and \etacts\ resolution
functions as the systematic uncertainty due to the mass-resolution
difference between data and MC.    
A systematic uncertainty on the mass accounts for the
different kinematics of two-photon signal and ISR \jpsi\ events.

The distortion of the resolution function due
to differences between the invariant mass distributions of the decay
products in data and MC produces negligible changes in the results.
We take as systematic uncertainty the changes in the resonance
parameters observed by including in the fit the effect of the efficiency
dependence on the invariant mass and on the decay dynamics.
%
The effect of the interference of the \etac\ signal with a possible
$J^{PC} = 0^{-+}$ contribution in the \gg\ background is considered. 
We model the mass distribution of the $J^{PC} = 0^{-+}$ background
component with the PDF describing combinatorial background.
The changes in the fitted signal yields are negligible. The changes
of the values of the \etac\ mass and width with respect to the nominal
results are $+1.2~\mevcc$
and $+0.2~\mev$ for \ks, and $+2.9~\mevcc$ and $+0.6~\mev$ for \kk.
We take these changes as estimates of systematic uncertainty due to
interference.  
The effect of the interference on the \etacts\ parameter values
cannot be determined due to the small signal to background ratio and the
smallness of the signal sample. We therefore do not include any
systematic uncertainty due to this effect for the \etacts.

Systematic uncertainties on the efficiency due to
tracking (0.2\% per track), \KS\
reconstruction (1.7\%), \piz\ reconstruction (3.0\%) and PID (0.5\% per
track) are obtained from auxiliary studies.
The statistical uncertainty of the efficiency parameterization is
estimated with simulated parameterized experiments. In each experiment, the
efficiency in each histogram bin and the coefficients of the functions
describing the dependence on  $\cos\theta_K$, $\cos\theta_{\pi\pi}$,
$\cos\Theta$, $\theta_\pi$ and $\Phi$ are varied within their
statistical uncertainties.  
We take as systematic uncertainty the width of the resulting yield
distribution. The fit bias is negligible. The small impact of the
presence of events falling in bins with zero efficiency is 
accounted for as an additional systematic uncertainty.

Using the efficiency-weighted yields of the \etac\ and \etacts\ resonances,
the number of peaking-background events, and 
$\calB(\KS\to\pip\pim)=(69.20\pm0.05)\%$~\cite{PDG}, we find the 
branching fraction ratios 
\begin{eqnarray}
 \frac{\calB(\etac\to\kk)}{\calB(\etac\to\ks)} =
 1.43\pm0.05\pm0.21,\label{eq:etacratio}\\
   \frac{\calB(\etacts\to\kk)}{\calB(\etacts\to\ks)} = 
   2.2\pm0.5\pm 0.5,\label{eq:etac2ratio}
\end{eqnarray}
where the first error is statistical and the second is
systematic. The uncertainty in the efficiency parameterization is the main
contribution to the systematic uncertainties and is equal to $0.17$
and $0.3$, in Eqs.~(\ref{eq:etacratio})
and~(\ref{eq:etac2ratio}), respectively. 
Using Eqs.~(\ref{eq:etacratio})--(\ref{eq:etac2ratio}),
$\calB(\etac\to K\overline{K}\pi)=(7.0\pm1.2)\%$ and $\calB(\etacts\to
K\overline{K} \pi) = (1.9\pm1.2)\%$~\cite{PDG} , and isospin
relations, we obtain 
$\calB(\etac\to\kk)= (3.3\pm0.8)\%$, and $\calB(\etacts\to\kk)=
(1.4\pm1.0)\%$, where we have summed in 
quadrature the statistical and systematic errors.

For each resonance and each final state, we compute the product
between the two-photon coupling $\Gamma_{\gg}$ and the resonance
branching fraction $\calB$ to the final state, using
$473.8~\invfb$ of data collected near the \FourS\ energy. 
The efficiency-weighted yields for the resonances, and the integrated
luminosity near the \FourS\ energy are used to obtain
$\Gamma_{\gg}\times\calB$ with the GamGam generator~\cite{BaBarZ}. 
The mass and width of the resonances are fixed to the values reported
in Table~\ref{tab:results}.
The uncertainties on the luminosity (1.1\%) and 
on the GamGam calculation (3\%)~\cite{BaBarZ} are included in the
systematic uncertainty of $\Gamma_{\gg}\times\calB$.
For the \ks\ decay mode, we give the results for the isospin-related 
$K\overline{K}\pi$ final state, taking into account
$\calB(\KS\to\pip\pim)=(69.20\pm0.05)\%$~\cite{PDG} and isospin
relations. 
For the $\chi_{c2}(2P)$, we compute $\Gamma_{\gg}\times\calB$ using the
fitted $\chi_{c2}(2P)$ yield, the integrated luminosity near the \FourS\
energy, and the average detection 
efficiency for the relevant process. The average detection efficiency
is equal to $13.9\%$ and $6.4\%$ for the \ks\ and \kk\ modes, 
respectively. 
The mass and width of the $\chi_{c2}(2P)$ resonance are fixed to the values
reported in~\cite{BaBarZ}. 
Since no significant $\chi_{c2}(2P)$ signal is observed, we determine a
Bayesian upper limit (UL) at 90\% confidence level (CL) on
$\Gamma_{\gg}\times\calB$, assuming a uniform prior probability
distribution. We compute the 
UL by finding the value of $\Gamma_{\gg}\times\calB$ below which lies
90\% of the total of the likelihood integral in the
$(\Gamma_{\gg}\times\calB)\geq0$ region. Systematic uncertainties are
taken into account in the UL calculation.  
Results for $\Gamma_{\gg}\times\calB$ for each resonance and final
state are reported in Table~\ref{tab:coupling}. 
The $\etac\to K\overline{K}\pi$ measurement is consistent 
with, but slightly more precise than, the PDG value~\cite{PDG}; 
the other entries are first measurements.

\begin{table}[htb]
\centering
\caption{Results for $\Gamma_{\gg}\times\calB$ for each resonance in
  $K\overline{K}\pi$ and \kk\ final states. The first uncertainty is
  statistical, the second systematic. Upper limits are computed at 90\%
  confidence level.}
\label{tab:coupling}
\begin{tabular}{lc}
\hline\hline
Process & $\Gamma_{\gg}\times\calB$~(\kev)\\
\hline
$\etac\to K\overline{K}\pi$ & $0.386\pm0.008 \pm 0.021$\\
$\chict\to K\overline{K}\pi$ & $(1.8\pm0.5 \pm0.2)\times 10^{-3}$ \\
$\etacts\to K\overline{K}\pi$ & $0.041\pm0.004 \pm0.006$ \\
$\chi_{c2}(2P)\to K\overline{K}\pi$ & $<2.1\times10^{-3}$ \\
\hline
$\etac\to \kk$ & $0.190\pm0.006 \pm0.028$ \\
$\chicz\to \kk$ & $0.026\pm0.004 \pm0.004$ \\
$\chict\to \kk$ & $(6.5\pm0.9 \pm1.5)\times10^{-3}$ \\
$\etacts\to \kk$ & $0.030\pm0.006\pm0.005$\\
$\chi_{c2}(2P)\to \kk$ & $<3.4\times10^{-3}$ \\
\hline\hline
\end{tabular}
\label{tab:coupling}

\end{table}

In conclusion, we report the first observation of \etac, \chicz\ and
\etacts\ decays to  \kk, with significances
(including systematic uncertainties) of $18\sigma$, $5.4\sigma$ and
$5.3\sigma$, respectively.  
This is the first observation of an exclusive  hadronic decay of
\etacts\ other than $K\overline{K}\pi$. We also report the first
evidence of \chict\ decays to \kk, with a significance
(including systematic uncertainties) of $4.0\sigma$, and have obtained
first measurements of the \chicz\ and \chict\ branching fractions to \kk.
The measurements reported in this paper are consistent with previous
\babar\ results~\cite{etactsBaBar,latestEtac}, and with world average
values~\cite{PDG}. 
The measurement of the \etacts\ mass and width 
in the the \ks\ decay supersedes the previous \babar\
measurement~\cite{etactsBaBar}.  
The measurement of the \etac\ mass and width 
in the the \ks\ decay does not supersede the previous \babar\
measurement~\cite{latestEtac}.  
The value of $\Gamma_{\gg}\times\calB$ is measured for each observed
resonance for both $K\overline{K}\pi$ and \kk\ decay modes. 
We provide an UL at 90\% CL on $\Gamma_{\gg}\times\calB$ for the
$\chi_{c2}(2P)$ 
resonance. 

We thank C.~P.~Shen and M.~Shepherd for useful discussions.
We are grateful for the excellent luminosity and machine conditions
provided by our \pep2\ colleagues, 
and for the substantial dedicated effort from
the computing organizations that support \babar.
The collaborating institutions wish to thank 
SLAC for its support and kind hospitality. 
This work is supported by
DOE
and NSF (USA),
NSERC (Canada),
CEA and
CNRS-IN2P3
(France),
BMBF and DFG
(Germany),
INFN (Italy),
FOM (The Netherlands),
NFR (Norway),
MES (Russia),
MICIIN (Spain),
STFC (United Kingdom). 
Individuals have received support from the
Marie Curie EIF (European Union),
the A.~P.~Sloan Foundation (USA)
and the Binational Science Foundation (USA-Israel).

%

\renewcommand{\baselinestretch}{1}


\begin{thebibliography}{99}

\bibitem{etacts}
S.~K~Choi \etal\ (Belle Collaboration), \jprl{89}, 102001 (2002);
K.~Abe \etal\ (Belle Collaboration), \jprl{89}, 142001 (2002).

\bibitem{CLEO}
D.~M.~Asner \etal\ (CLEO Collaboration), \jprl{92}, 142001 (2004).

\bibitem{etactsBaBar}
B.~Aubert \etal\ (\babar\ Collaboration), \jprl{92}, 142002 (2004).

\bibitem{etactsBaBarIncl}
B.~Aubert \etal\ (\babar\ Collaboration), \jprl{96}, 052002 (2006).


\bibitem{PDG}
Particle Data Group, K. Nakamura \etal, J. Phys. G {\bf 37}, 075021
(2010).  

\bibitem{Uehara}
S~Uehara \etal\ (Belle Collaboration), \epjc{53}, 1 (2008).

\bibitem{Ambrogiani}
M.~Ambrogiani \etal\ (E835 Collaboration), \jprd{64}, 052003 (2001).

\bibitem{massSplit}
S.~Godfrey and N.~Isgur, \jprd{32}, 189 (1985);
L.~P.~Fulcher, \jprd{44}, 2079 (1991); 
J.~Zeng \etal, \jprd{52}, 5229 (1995);
S.~N.~Gupta and J.~M.~Johnson, \jprd{53}, 312 (1996); 
D.~Ebert \etal, \jprd{67}, 014027 (2003); 
E.~Eichten \etal, \jprd{69}, 094019 (2004).

\bibitem{BelleX}
S.-K.~Choi \etal\ (Belle Collaboration), \jprl{91}, 262001 (2003).

\bibitem{Xconfirmation}
D.~E.~Acosta \etal\ (CDF Collaboration), \jprl{93}, 072001 (2004);
V.~M.~Abazov \etal\ (D0 Collaboration), \jprl{93}, 162002 (2004);
B.~Aubert \etal\ (\babar\ Collaboration), \jprd{71}, 071103 (2005).

\bibitem{newStates}
B.~Aubert \etal\ (\babar\ Collaboration), \jprl{95}, 142001 (2005);
T.~E.~Coan \etal\ (CLEO Collaboration), \jprl{96}, 162003 (2006);
C.~Z.~Yuan \etal\ (Belle Collaboration), \jprl{99}, 182004 (2007);
B.~Aubert \etal\ (\babar\ Collaboration), \jprl{98}, 212001 (2007);
X.~L.~Wang \etal\ (Belle Collaboration),  \jprl{99}, 142002 (2007);
S.-K.~Choi \etal\  (Belle Collaboration),  \jprl{94}, 182002 (2005);
B.~Aubert \etal\ (\babar\ Collaboration), \jprl{101}, 082001 (2008).

\bibitem{BelleZ}
S.~Uehara \etal\ (Belle Collaboration), \jprl{96}, 082003 (2006).

\bibitem{BaBarZ}  
B.~Aubert \etal\ (\babar\ Collaboration), \jprd{81}, 092003 (2010).

\bibitem{BABARNIM}
B.\ Aubert \etal\ (\babar\ Collaboration), \nima{479}, 1 (2002).

\bibitem{geant}
The \babar\ detector Monte Carlo simulation is based on GEANT4:
S. Agostinelli \etal, \nima{506}, 250 (2003).

\bibitem{Yang}
C.~N.~Yang, \pr{77}, 242 (1950).


\bibitem{BBW}
The power law tails are described by the function
$B(x)=\frac{(\Gamma_{(1,2)}/2)^{\beta_{(1,2)}}}{|x-x_0|^{\beta_{(1,2)}}+(\Gamma_{(1,2)}/2)^{\beta_{(1,2)}}}$,
  where $x_0$ is a parameter,
  $\Gamma_1$($\Gamma_2$) and $\beta_1$($\beta_2$) are used when 
  $x<x_0$ ($x>x_0$); see Ref.~\cite{latestEtac} for more information. 

\bibitem{CLEOprel}
K.~Gao, Ph.D. Thesis, University of Minnesota, 2008,
arXiv:0909.2818[hep-ex]; 
B.~Heltsley, ``New CLEO Results on Charmonium Transitions'', The
Sixth International Workshop on Heavy Quarkonia, Nara, Japan, 2008,
{\tt http://www-conf.kek.jp/qwg08/session1\_3/}\\{\tt heltsley.pdf}.

\bibitem{frodesen}
A.~G.~Frodesen \etal, \emph{Probability and Statistics in Particle
  Physics} (Universitetsforlaget, Bergen, Norway, 1979).

\bibitem{latestEtac}
J.~P.~Lees \etal\ (\babar\ Collaboration), \jprd{81}, 052010 (2010).
\end{thebibliography}
\end{document}